# All-dielectric Tailored-index Index-matched Arbitrary Waveguides based on Transformation Optics

Shahin Firuzi, Shengping Gong [*]
*School of Aerospace Engineering, Tsinghua University, 100084, Beijing, China*

**Abstract**

Transformation optics based on conformal mapping (CM) technique provides a practical approach to designing arbitrary-shaped all-dielectric devices. However, CM may result in large impractical refractive indices unachievable by current fabrication techniques. Furthermore, due to the low design flexibility of CM, the index-matching conditions at the ports of the device can not be realized as well. In this paper we introduce a method for tailoring the CM generated graded refractive index, in order to decrease its maximum value and apply index-matching, while maintaining the propagation path within the device unchanged. This technique is applied to two arbitrary index-matched waveguides. The full-wave simulation is performed to show the efficiency of the tailored-index index-matched waveguides for different relative wavelengths.

## 1. Introduction

Since the advent of transformation optics (TO) [1, 2], there have been many novel electromagnetic (EM) devices, including invisibility cloaks, flat lenses and reflectors, waveguide bends and couplers, and many other remarkable devices [3]. Metamaterials provide the necessary basis for realization of the TO designs. Among the large category of optical metamaterials [4], all-dielectric TO devices [5], are low-loss, provide a strong light confinement [6], and can be implemented with low cost CMOS-compatible fabrication techniques such as patterning air holes [7] or rods on a silicon on insulator (SOI) platform [8]. Conformal mapping (CM) and quasi-conformal mapping (QCM) [9], are the TO techniques used to design the arbitrary-shaped all-dielectric devices [10], including planar [11], and three-dimensional lenses [12–14], and different waveguides [15–20] which are utilized to bend, split, squeeze, and expand the EM waves. However, the CM/QCM techniques have a low design flexibility, which may result in large refractive indices unachievable by SOI technology.

---

[*] Corresponding Author.
 *E-mail addresses:* xiah16@mails.tsinghua.edu.cn (S. Firuzi), gongsp@tsinghua.edu.cn (S. Gong)

In this work, we introduce a modification method for altering the refractive index of all-dielectric devices, in order to confine it within the applicable range. This method can be also used for index-matching at the boundaries of the device, without requiring any extra index-matching layer. The design process and efficiency of this method are shown by two examples. We apply this method to designing two arbitrary index-matched waveguides with 180-degree and 90-degree bends, while the refractive indices of the input and output ports are matched to the neighboring media. The performance of the tailored-index devices are studied by performing a full-wave simulation, and the results are presented for different size parameters (inverse of relative wavelength). Furthermore, we also explain the dependency of the CM generated refractive index distribution to the design parameters, through describing the behavior of its minimum and maximum, as well as the average indices at the input and output ports. The current work provides the necessary basis for designing more complicated and radical all-dielectric structures which can be realized by the current fabrication technologies [21].

## 2. Theoretical basis

In this section, firstly we review the design process of a TO device by conformal mapping, and discuss the behavior of the resultant refractive index with respect to the design parameters. Then we present our proposed method for tailoring the resultant refractive index. The strict conformal mapping can be applied to design all-dielectric TO devices [25], by solving the Laplace's equation on the transformed space (TS) with the boundary conditions relating the TS to the original space (OS). The original space is considered to be a rectangle with the horizontal and vertical axes of $u$ and $v$. The transformed space is considered to be in the shape of the device structure with the $x$ and $y$ as its horizontal and vertical axes. Unlike the TS, the boundaries of the OS has a rectangular shape and can be described mathematically. Therefore, it is simpler to solve the Laplace's equation on the TS, with the boundary conditions given by the boundary equations of the OS. The OS and TS and the corresponding boundaries are shown in Fig. 1.

The Laplace's equation can be solved separately for $u$ and $v$ as,

$$u_{xx}^2 + u_{yy}^2 = 0, \quad v_{xx}^2 + v_{yy}^2 = 0 \tag{1}$$

where the subscripts $xx$ and $yy$ correspond to the second-order partial derivative with respect to $x$ and $y$, respectively. The boundary conditions can be divided into Dirichlet and Neumann conditions. For $u$ and $v$, the Dirichlet boundary conditions are defined as,

$$u|_{\Gamma_1} = 0, u|_{\Gamma_3} = c_1; \quad v|_{\Gamma_2} = 0, v|_{\Gamma_4} = c_2 \tag{2}$$

where $c_1$ and $c_2$ are positive constants. In every case, the Neumann conditions are applied to the remaining two boundary curves. As we will show later, the design of the refractive index distribution can be done by choosing the proper values for the constants $c_1$ and $c_2$. For a strict conformal mapping, however, $u$ and $v$ satisfy the Cauchy-Riemann conditions,

$$u_x = v_y, \quad u_y = -v_x \tag{3}$$

everywhere. Therefore, by having one of the variables, the other one is immediately known. It allows us to derive the refractive index distribution of the device by solving the Laplace's equation merely for either of $u$ or $v$ as,

$$n = n_0\sqrt{u_x^2 + u_y^2} = n_0\sqrt{v_x^2 + v_y^2} \tag{4}$$

where $n_0$ is the refractive index of the OS. In Eqs. (2) and (3), the constants $c_1$ and $c_2$ are related to each other through Cauchy-Riemann equations. Therefore, by considering a strict conformal map, by having one of these constants, the other one is also known.

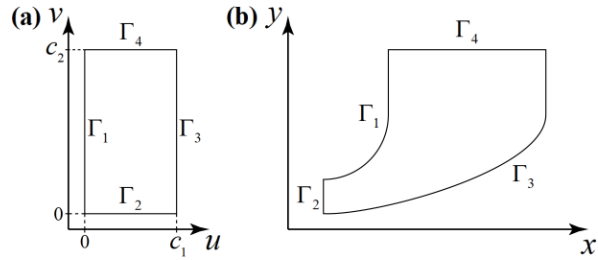

**Fig. 1 OS and TS and the corresponding axes: (a) OS; (b) TS.**

In this paper, the design of the refractive index distribution is done by considering a CM design through solving the Laplace's equation for $u$ while adjusting the constant $c_1$. Here, we describe the dependency of the refractive index distribution to the values of $c_1$, simply by considering its minimum ($n_{\min}$) and maximum ($n_{\max}$) values, as well as its average values at the input ($n_{\text{in}}$) and output ($n_{\text{out}}$) ports. The minimum and maximum indices throughout the device are always positive, regardless of how small $c_1$ is chosen. However, scaling the $c_1$ by a certain rate, results in the whole resultant refractive index distribution to be scaled by the same rate. This means, by considering a uniform $n_0$, we don't have a full control over defining the $n_{\min}$, $n_{\max}$, $n_{\text{in}}$, and $n_{\text{out}}$. Therefore, by tuning $c_1$, only one of these variables can be adjusted, and the other variables are defined accordingly. Normally, for designing an all-dielectric device, $c_1$ is tuned in such a way as to achieve $n_{\min}=1$. Accordingly, the value of $n_{\max}$ is determined regarding the shape of the device, while $n_{\text{in}}$ and $n_{\text{out}}$ are inversely proportional to the widths of the ports. By considering the lack of control over defining

$n_{max}$, $n_{in}$, and $n_{out}$, the conformal mapping may result in large values of $n_{max}$ unachievable by the fabrication techniques of all-dielectric structures [21], and furthermore, the index-matching at the ports also can't be done by a simple use of conformal mapping techniques.

Here we show how the refractive index distribution can be modified to achieve two main goals; decreasing the maximum refractive index, and matching the indices of the input and output ports to the neighboring media. It is a known fact that the refractive index gradient along the propagation direction, doesn't affect the propagation direction of the waves. In this study, the propagation direction and wavefronts in OS are considered to be along $v$ and $u$, respectively. Therefore, applying any arbitrary refractive index gradient along $v$ doesn't affect the propagation direction. This feature provides a great design flexibility through modification of refractive index along $v$, in order to alter the CM generated refractive index in a desired manner. Two coefficients of $\Phi_1(v)$ and $\Phi_2(v)$ are utilized to define the tailored index distribution as,

$$n' = \Phi_1(v) \cdot \Phi_2(v) \cdot n \tag{5}$$

where $\Phi_1$ and $\Phi_2$ are used for confining the refractive index and index-matching, respectively.

After achieving the refractive index distribution $n(x,y)$ by CM, the index tailoring and matching is applied by five steps as below.

1. The CM generated refractive index distribution $n(x,y)$, which is given in TS by $x$ and $y$, will be transferred into OS numerically, so to be expressed by $u$ and $v$ as $n_T(u,v)$.

2. The minimum and maximum values of $n_T(u,v)$ over $u$, is derived as,

$$\begin{cases} n_{T,\max}(v) = \max\{n_T(u,v) : 0 \leq u \leq c_1\} \\ n_{T,\min}(v) = \min\{n_T(u,v) : 0 \leq u \leq c_1\} \end{cases} \tag{6}$$

3. The coefficient $\Phi_1(v)$ is then given corresponding to $n_{T,\max}(v)$ and $n_{T,\min}(v)$, in order to restrain them within a desired range.

4. The coefficient $\Phi_2(v)$ is then given according to $\Phi_1(v_{in}).n_{in}$ and $\Phi_1(v_{out}).n_{out}$, in order to match the refractive indices of the both ports to their neighboring media. The notations $v_{in}$ and $v_{out}$ correspond to the values of $v$ at the input and output ports, respectively.

5. Lastly, the tailored matched index $n'_T(u,v) = \Phi_1(v).\Phi_2(v).n_T(u,v)$ is achieved and transferred back to the TS as $n'(x,y)$.

## 3. Simulations

In this section, we use two examples to show how tailoring the refractive index and index-matching can be done. In the first case, an arbitrary waveguide with a 180-degree bend and different port widths is considered, while the refractive indices of the ports are matched to the vacuum. In the second case, we consider a waveguide with a 90-degree bend with different port widths as well as different refractive indices of the neighboring media at each port. In both cases we compare the performance of the tailored-index devices with the original device resulted from conformal mapping, and show the efficiency of the device for different size parameters ($R$).

### 3.1. First example

The structure of the waveguide is shown in Fig. 2(a). It is composed of a bend and two rectangular guides. The conformal mapping results in a refractive index distribution which is shown in Fig. 2(b). According to an intrinsic characteristic of conformal mapping, the uniformity of the resultant index on the input and output boundaries depends on the sharpness of the bend as well as the lengths of two rectangular parts of the waveguide shown in Fig. 2(a). As it can be seen from Fig. 2(b), the refractive index becomes more uniform by moving along the straight rectangular parts of the waveguide towards the ports. Therefore, in order to achieve a desirably uniform index at the ports, adding a straight part to the waveguide after the bend is necessary. We need to mention that these rectangular straight guides are a part of the waveguide on which the CM is applied. For the simplicity of the design, the parameter $c_1$ is defined is such a way as to achieve $n_{min}=1$. However, it results in very large and impractical maximum index of $n_{max}=10.64$. By applying the proposed tailoring method, as it can be seen in Fig. 2(c), the maximum index can be dramatically reduced to $n'_{max}=2.5$.

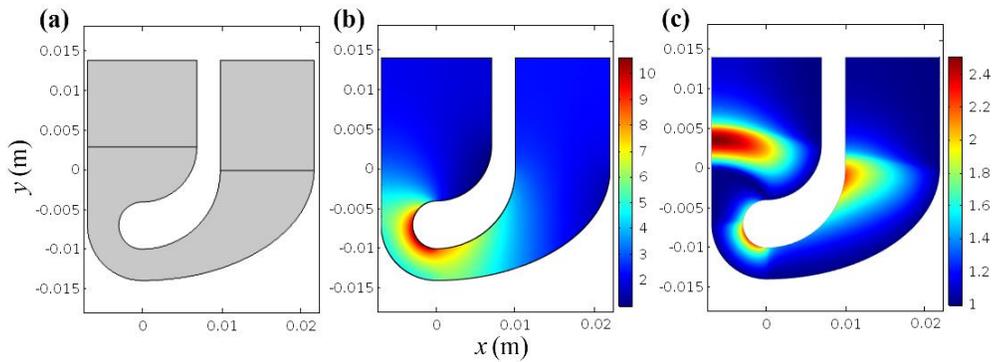

**Fig. 2 First example: (a) the structure; (b) the CM index; (c) the tailored index of the waveguide.**

Here we show the modification process in detail. As the first step of the modification process, the index $n(x,y)$ is transferred into the OS as $n_T(u,v)$, which is shown in Fig. 3(a). Then, according to the values of $n_T(u,v)$, the $n_{T,\max}(v)$ and $n_{T,\min}(v)$ are derived. In this design, our aim is to minimize $n'_{T,\max}(v)$ while keeping $n'_{T,\min}(v)=1$. Therefore, the coefficient $\Phi_1(v)$ is derived as $\Phi_1(v)=1/n_{T,\min}(v)$, which is shown in Fig. 3(b). Since the condition of index-matching has been already achieved by the tailoring process ($n'_{in}=n'_{out}=1$), then the second coefficient is consider as $\Phi_2(v)=1$. Accordingly, the modified refractive index which is shown in Fig. 3(c), is obtained by $n'_T(u,v)=\Phi_1(v).\Phi_2(v).n_T(u,v)$. It can be seen that for the modified index we have $n'_{T,\min}(v)=1$ for all $v$, even at the ports of the device which satisfies the condition of the index-matching to the vacuum. The tailored index is then transferred back to the TS, which is shown in Fig. 2(c).

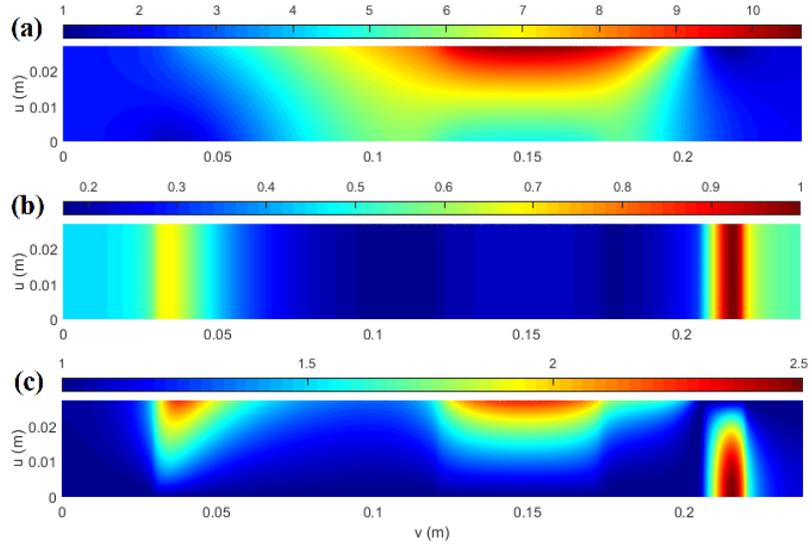

**Fig. 3  First example: (a) CM generated index; (b) the values of $\Phi_1$; (c) tailored index distribution in OS.**

In order to show the performance of the modified index and compare it with the CM generated index, we perform a full-wave simulation by considering a range of size parameters. The size parameter is given by $R=(c_1.\min\{\Phi_1.\Phi_2\})/\lambda_0$, where $\lambda_0$ is the wavelength in the vacuum. The results of the simulation is shown in Fig. 4. The performance of the CM index is shown in Fig. 4(a), where the results of the modified index are shown in Fig. 1(b) to Fig. 1(f). It can be seen that the device has a very good performance at larger size parameters. However, at smaller size parameters the efficiency of the device decreases. We can describe this by explaining the ability of the device in squeezing the wave (beam confinement). By this example, the role of the proposed index-tailoring method becomes clearer. In the device with modified index we sacrifice a part of the squeezing ability of the device for achieving a lower maximum index, while keeping the propagation direction identical to the original CM generated index. Therefore, the confinement ability of the tailored-index waveguide decreases by decreasing the size parameter. The efficiency of the power transmission of the index-tailored device is shown in Fig. (8). It can be seen that the efficiency of the device is about 97.1 % for R=10, and above 80 % for R>4.

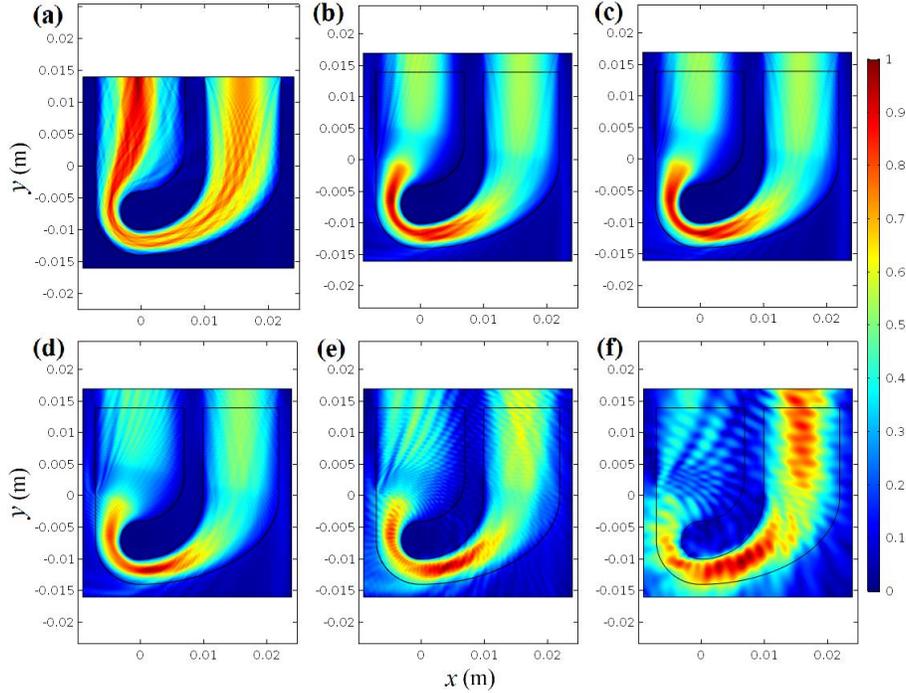

**Fig. 4** First example, the results of the full-wave simulation: (a) device with CM generated index; (b), (c), (d), (e), and (f) correspond to the size parameters of 10, 8.5, 5.5, 2.5, and 1, respectively.

*3.2. Second example*

The structure of the 90-degree waveguide is shown in Fig. 5a. In this case, the refractive indices of the neighboring media at the wide (input) and narrow (output) ports are considered to be 1.5 and 1.2, respectively. The refractive index distribution of the device resulted from conformal mapping is shown in Fig. 5b. It can be seen that the maximum refractive index is $n_{max}$=8.14, and the average refractive indices of the input and output ports are $n_{in}$=1.83 and $n_{out}$=7.31, respectively. By applying the tailoring process, the refractive index variation is minimized and the index matching is applied to the ports. The resulting refractive index distribution is shown in Fig. 5c. It can be seen that the maximum value of the tailored index is $n'_{max}$=2.45, and the average port indices are matched to the required values.

The CM generated index distribution as presented in the OS, $n_T(u,v)$, is shown in Fig. 6a. The coefficient $\Phi_1$ is then derived as $\Phi_1(v)=1/n_{T,min}(v)$, which is shown in Fig. 6b. Then the tailored index, before index-matching, is derived as $\Phi_1(v)$. $n_T(u,v)$ and shown in Fig. 6c. It can be seen that at this stage, the minimum index and the average port indices are equal to unity. In order to apply index-matching, the coefficient $\Phi_2(v)$ is derived according to the values of $\Phi_1(v_{in}).n_{in}$ and $\Phi_1(v_{out}).n_{out}$, where in this example $v_{in}=0.0293$ (m) and $v_{out}=0$. Since in this stage we have achieved $\Phi_1(v_{in}).n_{in}=1$ and $\Phi_1(v_{out}).n_{out}=1$, we need to define $\Phi_2(v)$ in such a way as to achieve $\Phi_1(v_{in}).\Phi_2(v_{in}).n_{in}=1.5$ and $\Phi_1(v_{out}).\Phi_2(v_{out}).n_{out}=1.2$. Therefore, we define $\Phi_2(v_{in})=1.5$ and $\Phi_2(v_{out})=1.2$. The values of $\Phi_2$ for other $v$ except $v_{in}$ and $v_{out}$ is defined arbitrarily in such a way as to decrease its effect on the resultant refractive index, especially on the minimum and maximum index. The coefficient $\Phi_2$ is shown in Fig. 6d. The tailored matched index is then derived as $n'_T(u,v)= \Phi_1(v).\Phi_2(v).n_T(u,v)$, which is shown in Fig. 6e. It's obvious that $n'_T(u,v)$ satisfies the desirable index-matching condition, while minimizing the maximum index and keeping the minimum index above unity. However, it can be seen that the resultant index is not completely uniform at the input port, which is caused by the non-uniformity of the original CM generated index at the input port. As we described before, it can be improved by increasing the length of the straight waveguide at the input port. At the end, the modified index is transferred back to the OS, which is shown in Fig. 5(c).

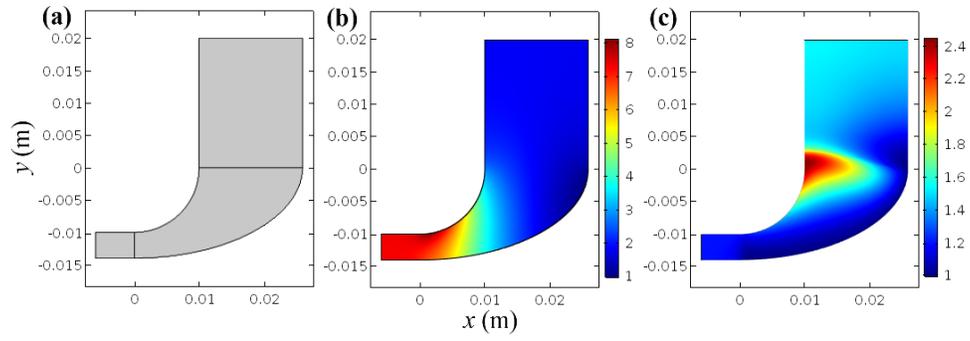

**Fig. 5** Second example: (a) the structure; (b) the CM index; (c) the tailored index of the waveguide.

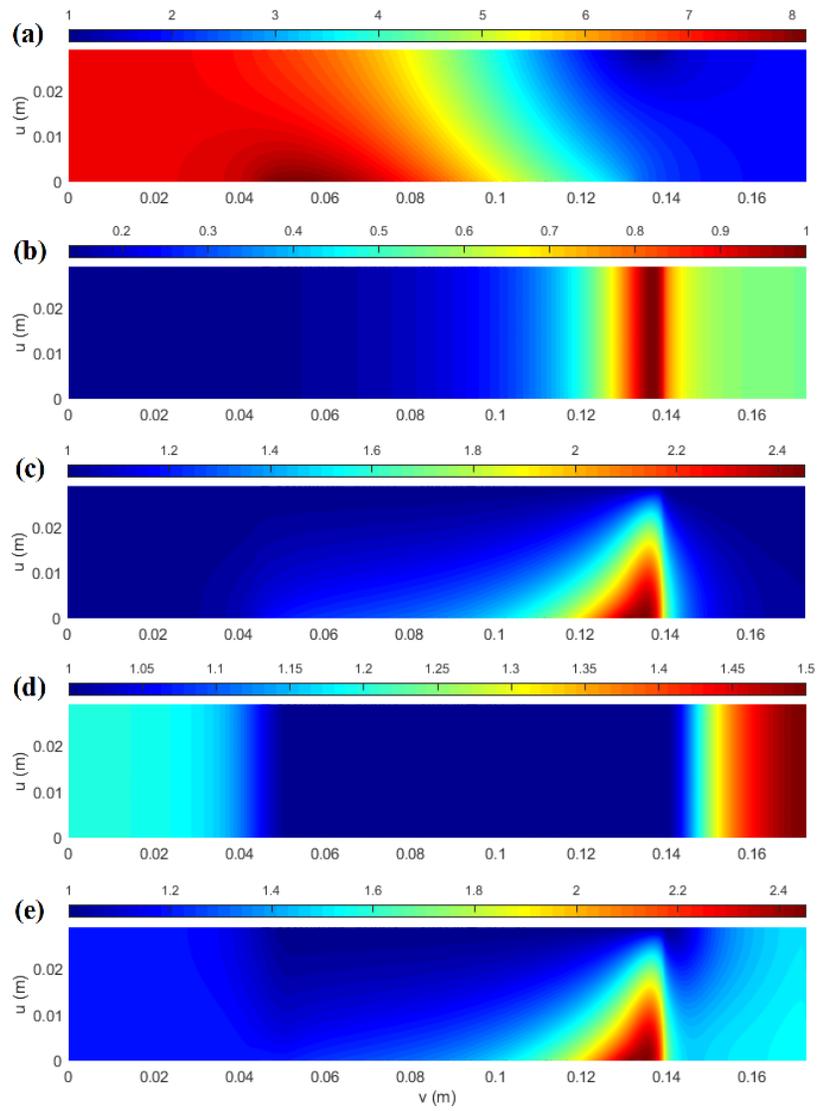

**Fig. 6** Second example: (a) the CM generated index; (b) the values of $\Phi_1$; (c) the tailored index distribution; (d) the values of $\Phi_2$; (e) the tailored matched index distribution in OS.

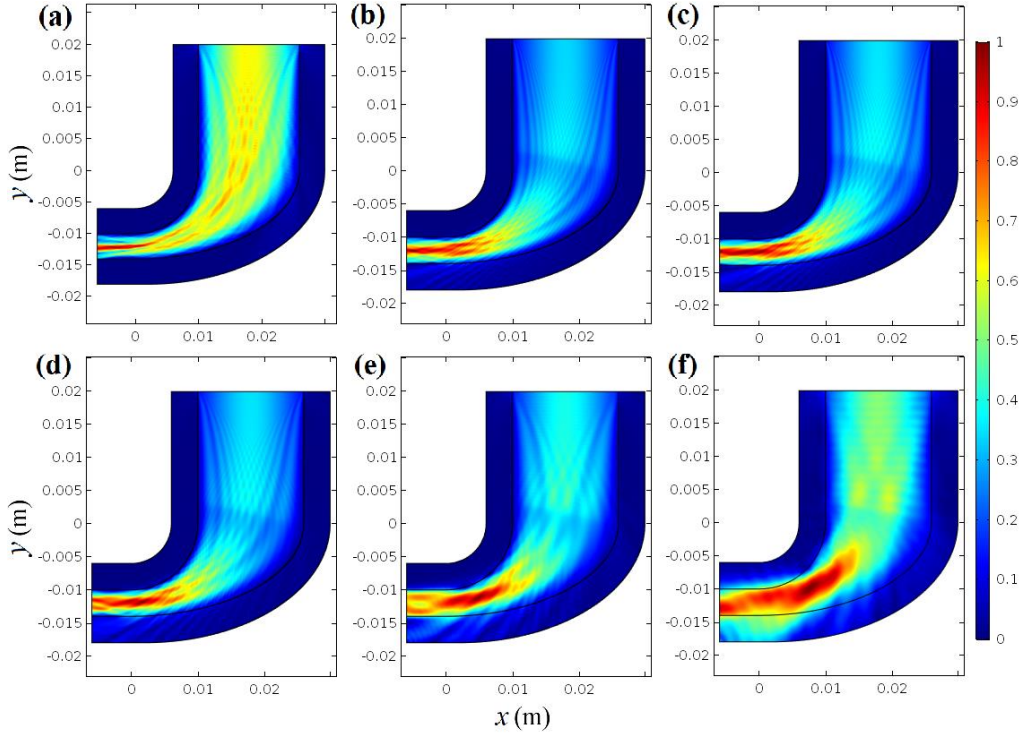

**Fig. 7** Second example, the results of the full-wave simulation: (a) device with CM generated index; (b), (c), (d), (e), and (f) correspond to the size parameters of 10, 8.5, 5.5, 2.5, and 1, respectively.

The performance of the tailored-index device is compared to the original device by a full-wave simulation for different size parameters similar to the first example. The result of the simulation of the original device has been shown in Fig. 7(a), and the results of the tailored-index index-matched device are shown in Fig. 7(b) to Fig. 7(f). The power transmission efficiency of the waveguide for different size parameters has been measured and shown in Fig. 8. The efficiency of the waveguide at R=10 is computed as 98.6 %, and it has an efficiency of greater than 80 % for R>2.

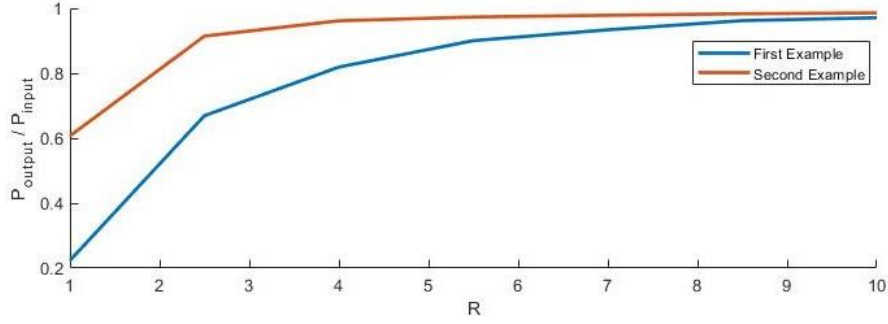

**Fig. 8  The efficiency of the waveguides in delivering the energy for different size parameters $R = (c_1 . \min\{\Phi_1 . \Phi_2\})/\lambda_0$.**

## 4. Conclusions

We have presented a method for designing practical all-dielectric devices by modifying the gradient refractive index generated through conformal mapping. By applying this technique, the maximum value of the refractive index can be reduced dramatically, and the index-matching can be applied within the boundaries of the device. This method reduces the maximum index by lowering the ability of the device in squeezing the waves, while keeping the wave-guiding abilities identical to the original device. This indicates that the devices designed through this method have a lower performance for large wavelengths, while perform very well for shorter wavelengths. The full-wave simulation shows an efficiency of 97.1 % and 98.6 % at a size parameter of 10 for a 180-degree and a 90-degree waveguide bend, respectively. Furthermore, an efficiency of more than 80 % can be achieved for the 180-degree and 90-degree waveguides at the size parameters greater than 4 and 2, respectively. However, the efficiency of the devices may be increased further by relaxing the desired value of the maximum index. The presented method provides a solution for designing practical index-matched sharp waveguide bends and couplers with a higher flexibility.

**Conflict of interest statement**

The authors declare that there is no conflict of interests regarding the publication of this article.